\documentclass[oneside,10pt]{article}
\usepackage{epsfig}
\usepackage{graphicx}
\usepackage{epstopdf}
\def\sub#1{\vspace{-3mm}\subsection*{\it {#1}}\vspace{-3mm}}  
\def\bf{\textbf}

\def\ni{\noindent}

\def\ni{\noindent}

\newcommand{\bea}{\begin{eqnarray}}
\newcommand{\eea}{\end{eqnarray} \noindent}

\def\ni{\noindent}
\newcommand{\be}{\begin{equation}}
\newcommand{\ee}{\end{equation} \noindent}

\begin{document}
\pagestyle{plain}

\begin{flushright}
\end{flushright}

 \begin{center} 

\large \bfseries
Does gravity operate between galaxies?  Observational evidence re-examined

\normalfont

Francis.J.M. Farley \\

\normalsize
\end{center}
School of Engineering and the Environment, Southampton University, Highfield, Southampton SO17 1BJ, England.
(Email: F.Farley @ soton.ac.uk)



 \baselineskip = 7mm    

\sub{Summary}
The redshifts and luminosities of Type 1A supernovae are conventionally fitted with the current paradigm, which holds that the galaxies are locally stationary in an expanding metric.  The fit fails unless the expansion is accelerating; driven perhaps by "dark energy".  

\ni Is the recession of the galaxies slowed down by gravity or speeded up by some repulsive force?  

\ni To shed light on this question the redshifts and apparent magnitudes of type 1A supernovae are re-analysed in a cartesian frame of reference omitting gravitational effects. The redshift is ascribed to the relativistic Doppler effect which gives the recession velocity when the light was emitted; if this has not changed, the distance reached and the luminosity follow immediately.  This simple concept fits the observations surprisingly well with Hubble constant $H_0 = 62.9 \pm 0.3  \ km/s/Mpc$.  It appears that the galaxies recede at unchanging velocities, so on the largest scale there is no significant intergalactic force.  Reasons for the apparent absence of an intergalactic force are discussed.




 \sub{Introduction}

It is natural to assume that the inverse square law of gravity is valid over any distance and therefore operates between galaxies.  Ideally this should be verified by observation.  Now very accurate data, (Kowalski  \emph{et al} 2008), (Hicken  \emph{et al} 2009), painstakingly collected, on the apparent luminosity and redshift of type 1A supernovae enables us to measure the recession of the galaxies over great distances and to look for the effect of gravity.  If they are attracted to each other, the recession should be slowing down.

According to Hubble, the galaxies are receding from earth with velocities proportional to distance.  In Minkowski space the world lines of the galaxies, extrapolated back in time, would all diverge from a common point on the time axis, referred to for convenience as time zero or the big bang.  This provides a convenient origin of coordinates.  The view from another galaxy (galaxy B) is obtained by rotating the coordinates about the origin (Lorentz transformation) so that the world line of galaxy B becomes the new time axis.  The world lines of all galaxies, unaffected by this rotation, would still diverge from the origin.  The observer on galaxy B would therefore see the Hubble law, although the Hubble constant would be modified by the time dilation between earth and galaxy B.  Which frame should we choose to represnt the big bang universe?

In the ideal frame the big bang would be isotropic with an equal number of galaxies seen in all directions.  This is more or less true for the observer on earth; perhaps because distant, fast moving galaxies are too faint and too redshifted to be detected and this threshold is the same in all directions.  For an observer on galaxy B, new galaxies will come into view and others will drop out; the visible galaxies may again appear isotropic.  If the universe looks isotropic in all frames, with the same Hubble law, then any one of them could be nominated with equal validity as the centre of the big bang.   In this context it may be reasonable in calculations to adopt a geocentric frame of reference, not implying that this frame is in some sense privileged, but instead that it is typical of a frame moving with any galaxy.  However the cosmic microwave background (CMB) does define a preferred frame and the earth's velocity relative to the CMB is quite small.

For the present study we need make no assumptions about cosmology.  The recession of the galaxies is observed from earth.  Have their velocities, relative to earth, remained constant; or are they slowing down?  For this purpose it is convenient to use a frame of reference with the earth stationary at the origin.  We analyse the kinematics of the supernovae, completely ignoring gravitational effects which might accidentally prejudice the inquiry.  We therefore eschew the metrics of general relativity and adopt instead a simple static cartesian frame of reference with no curvature and use special relativity.  Whether this cartesian frame can be extended to astronomical distances should be decided by the data.  This is also the frame in which the known laws of physics have been established.  In particular, the relativistic Doppler shift of photon energy (wavelength) and the independence of the velocity of light from source motion, (Alv\"ager \emph{et al} 1964), have both been accurately checked for velocities up to $0.99975 \times c$.  Time dilation in a circular orbit has been established at high accuracy for elementary particles (Bailey, Farley and Picasso 1968) and less accurately but with inclusion of the gravitational redshift by H\"afele and Keating (1972).  The latter is an essential input to contemporary satellite navigation systems (Ashby 2002).

In this picture the galaxies, (or the primeval matter from which they evolved), were all close together 
at some moment referred to as time zero and  diverged from a more or less common point with various velocities $v$.  The distance reached at any time is then proportional to $v$ and Hubble's's law follows naturally.  The question is, does each galaxy move out at constant velocity relative to earth or is it gradually slowed down by gravity.

Many studies of supernovae of type 1A (SN1A) have been reported.  It is established that with few exceptions their intrinsic luminosity is the same, so they can be used as standard candles.  Their apparent magnitude $M$ is then a measure of their distance from earth when the light was emitted.  The redshift $z$ indicates the velocity of recession from earth.  The  Hubble plot of $M$ vs $z$ then gives information about the recession of the galaxies.  A recent compilation of the data on SN1A, with many references, has been given by Kowalski \emph{et al} (2008) and further data has been added by Hicken  \emph{et al} (2009). 

Our approach is to calculate the apparent magnitude of the SN1A supernova as a function of its redshift, on the assumption that the galaxies recede at unchanging velocities.  In special relativity the redshift of the supernova determines its velocity when the light was emitted.  If the surrounding galaxy had moved uniformly at this speed since time zero, its distance at the moment of light emission is easily calculated.  The inverse square law then determines its apparent magnitude.    We then compare this theory with the data.  The supernova magnitude depends on its mean velocity since time zero, while the redshift is fixed by the final velocity.  Therefore if the galaxies are slowing down, or perhaps speeding up, the data will not agree with the prediction.

\sub{Theory}
In this picture the light emitted by a moving supernova is transformed to the rest frame of the observer, assumed stationary, using the Lorentz transformation.  The Doppler shift and solid angle adjustment occur immediately and the photon then propagates unchanged to the receiver.  The luminosity varies as $1/r^2$ where $r$ is the distance from earth when the light was emitted.  This will be a function of the unchanging velocity $v=\beta  c$ of the galaxy relative to earth.  The reported magnitudes of SN1A are taken to represent the energy flux received at earth at the peak of the luminosity curve (Guy  \emph{et al} 2005).  In applying the Lorentz transformation of special relativity the following factors must be included:

\ni 1) The redshift $z$ is given by $\lambda = \lambda _0 \ (1 + z)$ with
\be
1 + z = \sqrt { \frac{1+\beta}{1-\beta} }     \label{z}
\ee
where the velocity of recession from earth is $\beta c$.  This implies
\be
\beta = \frac{2z+z^2}{2+2z+z^2}   \label{beta}
\ee
\ni 2)  Transformation of solid angle, Jacobian.  For light emitted in the backward direction by a moving source, the transverse momentum is not affected by the motion while the longitudinal momentum is reduced by the factor $1/(1+z)$.  As a result the solid angle of the emitted light is increased by $(1+z)^2$.  The luminosity is decreased by the factor $1 / (1+z)^2$.

\ni 3) The energy of the photons emitted backwards is reduced by the factor $1 / (1+z)$.  As stellar magnitude is calculated from the energy flux, there is a corresponding reduction in the luminosity.

\ni 4) Time transformation.  For an observer at rest, everything in the frame moving away from him is apparently slowed down: time is stretched by the factor $(1+z)$.  The redshift noted above is an example of this.  It applies also to the temporal evolution of the supernova luminosity, which has been carefully verified by Goldhaber \emph{et al} (2001) and Blondin \emph{et al} (2008).  As the received light is spread out over a longer time, the peak of the luminosity curve must be decreased by the factor $1 / (1+z)$.

\begin{figure}
\begin{center}
\resizebox{8cm}{!}{ \includegraphics{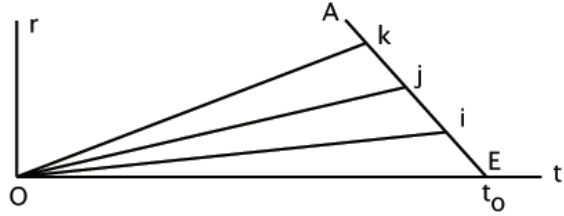}}
\caption{Supernova distance $r$ vs time with light path to earth AE}
\label{f1}
\end{center}
\end{figure}

\ni 5) The time for the light to travel from the SN to earth must be considered; distant SN emitted their light long before it is received.  This is illustrated in Figure 1, which shows the distance $r$ of the SN from earth versus the time $t$ from time zero.  The SN diverge from time zero along lines \emph{Oi, Oj, Ok} at various velocities $\beta c$, while the light received at earth at time $t_0$ (now) follows the line \emph{AE}.  The intersection points \emph{i, j, k} determine the distance $r$ at the moment the light was emitted.  It follows that
\be
r = \frac{\beta c t_0}{1+\beta}  \label{r}
\ee
The received luminosity is proportional to
\be
\frac{1}{r^2} \propto \frac{(1+\beta) ^2}{\beta ^2}  \label{td}
\ee

\ni Combining all these factors and using eqn (\ref{beta}), the observed luminosity $L$ is given by
\be
L = \frac{L_0}{c^2 t_0^2} \frac{(1+\beta) ^2}{\beta ^2}  \frac{1}{(1+z)^4} = \frac{L_0}{c^2 t_0^2}\frac{4}{(2z+z^2)^2}   \label{LU}
\ee
$t_0$ is the Hubble time.  Therefore the observed SN magnitudes at redshift $z$ should be
\be
M = -2.5\  log_{10} (L) = M_0 + 5\  log_{10}(z+z^2/2)  \label{M}
\ee

\ni $M_0$ is related to the Hubble constant $H_0$ as follows.  The "constitution" data set tabulated by Hicken \emph{et al} (2009) gives the apparent magnitude $m$ of the supernova relative to its absolute magnitude (taken as zero) corresponding to distance $r_0 = 10\ pc$.  If the distance to the supernova is $r$ 
\be
m = 5 \ log_{10} (r/r_0)  \label{snm}
\ee
For small distances, small velocities, $z = \beta$ and according to the Hubble law the velocity  $\beta c = H_0 r$, so 
$r = c z / H_0$ and (\ref{snm}) becomes
\be
m = 5\  log_{10} (c / H_0 r_0) + 5\  log_{10} (z)  \label{snm2}
\ee
Comparing with (\ref{M}) at small $z$,
\be
M_0 = 5\  log_{10} (c / H_0 r_0)   \label{M0}
\ee

\sub{Fitting the data}

Equation (\ref{M}) is plotted in Figure 2, together with the ''constitution data set'' of supernova luminosity versus redshift recommended by Hicken \emph{et al} (2009).  A logarithmic scale is used for $z$ to spread out the data more clearly.  In the theory the only free parameter is $M_0$, which moves the theoretical curve vertically.  To give the best fit to the data, its value is adjusted using the weighted least squares method, including the observational uncertainties tabulated by Hicken et al, with the result 
\be
M_0 = 43.394 \pm 0.010  \label{M0value}
\ee
The $\chi ^2$ for the fit is 497 for 396 degrees of freedom, indicating slightly more variability in the data than expected from the quoted errors.  The error in $M_0$ has been increased to allow for this.
There are no free parameters to manipulate the shape of the curve nor its slope: apparently none are needed.
To compare the theory more exactly with the data, Figure 3 shows the residuals with observational uncertainties on a magnified scale.   The fit of the theory to the data is excellent.  It is remarkable that the simple algebraic formula (\ref{M}) with only one free parameter can fit the data so closely.  This seems to justify our approach.

If the galaxies are slowing down, the redshift $z$ is set by the final velocity, while the SN magnitude is set by the distance reached which depends on the mean velocity, which would be higher; the magnitude would be too large and the points should lie above the line in Figure 2.  The slower moving galaxies would be affected most by gravity and show the largest $\Delta M/M$:  first, they have less kinetic energy, and secondly being closer to earth, the light was emitted later so there was more time for gravity to act.  Therefore the residuals at small $M$ would be positive, while the data at large $M$ would be less affected.  Conversely, if the galaxies are speeding up, the residuals at small $M$ would be negative.    There is no clear evidence in Figure 3 for either effect.

From Figure 3, a discrepancy of more than $0.2$ magnitudes is excluded, corresponding to a change of $\approx 10\%$ in the mean recession velocity compared to the final velocity, or a $20\%$ final change for the slowest galaxies since time zero.
Within these limits it appears that the galaxies are receding at unchanging velocities.  They are neither slowing down nor speeding up.  On the largest scale, there appears to be no force decelerating or accelerating a galaxy, therefore no indication of gravity.  This justifies, \emph{a posteriori} our use of cartesian coordinates with zero curvature with light propagating in straight lines: the analysis is self consistent.

Using (\ref{M0}) and (\ref{M0value}) the fitted value of $M_0$ gives the Hubble constant
\be
H_0 = 62.9 \pm 0.3  \ km/s/Mpc  \label{H}
\ee
This is in excellent agreement with the result of a 15 year study  using Cepheid variables to calibrate the cosmic distance scale by Sandage et al (2006) who recommend $H_0 = 62.3 \pm 1.3\ (statistical) \pm 5.0\   (systematic)\ km/s/Mpc $.
In contrast, Riess et al (2009) have applied a cosmological model to derive  $H_0 = 74 \pm 4 \ km/s/Mpc$: but  Sandage et al (\emph{loc. cit.}) have given reasons for discarding values in this range.

In summary, the assumption that the galaxies are receding from earth at unchanging velocities in a fixed metric with no curvature gives a good account of the data and a reasonable value of the Hubble constant in support of Sandage et al.

\sub{Discussion}
On the largest scale, the net force accelerating or decelerating the galaxies is apparently zero.  Each recedes from earth with unchanging velocity following Newton's first law. There are several possibilities.

1) The galaxies are attracted to each other by gravity, but there is another repulsive force which cancels it, perhaps called ''dark energy'' or ''the cosmological constant''.  We then need to explain the origin of dark energy and why it cancels gravity so exactly.  This cancellation was apparently valid when the galaxies were closer together, so dark energy would have to follow the same inverse square law as gravity.  Why does this repulsive force not show up inside the galaxies?  While the observations are indicating zero force, the evidence for a new hitherto unknown force is perhaps not compelling.

2) Gravity falls off at large distances faster than the inverse square law.  This would also arise if gravity is mediated by gravitons (Feynman 1995) and the graviton had a very small, but finite mass. 

3) Assume that the universe is isotropic on the largest scale and infinite.  Then the gravitational potential will also be isotropic and the same everywhere.  With no gradient in potential, there will be no force of gravity to act upon the galaxies.  This argument applies not only to gravity, but to all force fields; whatever they are, they can give no net force.  Force is a vector and must have a direction: in an isotropic universe no direction is preferred, so no force.  This of course does not exclude attractive or repulsive forces on a local scale, wherever the distribution of matter is no longer isotropic; as within galaxies or galactic clusters, etc.  (For this argument to apply, the universe may not need to be infinite but large compared to the accessible volume).

A major attraction of this hypothesis is that it explains the flatness problem, which has been called ''one of the great mysteries of cosmology'', why the metric favoured by experts on general relativity is flat with no curvature, implying no general acceleration or deceleration, (Feynman 1995-2),  (Liddle 2003).  In an infinite isotropic universe there can be no general acceleration or deceleration, so the metric is automatically flat and remains flat at all times.  In this picture, matter at time zero was not localised, but appeared everywhere at the same time; and immediately after, the universe was infinite, isotropic and expanding.  With the Hubble law applying locally everywhere, the velocity addition theorem of special relativity implies that superluminal velocities are not involved.  Very distant galaxies recede at velocities asymptotic to $c$, with large redshifts and low apparent luminosity as suggested by eqn (\ref{LU}) above.

Other fits to the SN1A data  have been reviewed with many references by Copeland  \emph{et al} (2006).  Starting with different assumptions they arrive at different conclusions, which now appear to be model dependent.

\sub{Acknowledgements}
I wish to thank Martin Rees and John Field for helpful discussions and Timothy Farley and Neville Moray for advice on statistics.

\newpage

\sub{\bf{References}}
\ni  Alv\"ager, T., Bailey, J.M., Farley, F.J.M., Kjellman, J. \& Wallin, I.  1964 \emph{Physics Letters} \bf{12}, 260; 
 \emph{Arkiv f\"{o}r Fysik}, \bf{31}, 145 (1966)
 \newline\ni Ashby, N. 2002 \emph{Physics Today}, May 2002, pp. 41-47
  \newline\ni Bailey, J.M., Farley, F.J.M. \&  Picasso, E. 1968 \emph{Nature }\bf{217}, 17 (1968);  Bailey  J. \emph{et al} 1977 \emph{Nature} \bf{268}, 301
 \newline\ni Blondin, S. \emph{et al}, 2008 \emph{Astro. J.} \bf{682}, 724
 \newline\ni Copeland, E. J., Sami,  M. \& Tsugikawa, S. 2006 \emph{Int. J. Modern Physics} \bf{D15},1753
 \newline\ni Feynman, R.P. 1995 \emph{Lectures on Gravitation}, Penguin, London 
 \newline\ni Feynman, R.P. 1995-2 \emph{loc.cit} p.10
 \newline\ni Goldhaber, G. \emph{et al}, 2001 \emph{Astro. J.} \bf{558}, 359 
 \newline\ni Guy, J., Astier, P., Nobili,S., Regnault, N. \& Pain,R. 2005 \emph{Astronomy \& Astrophysics} \bf{443}, 781
 \newline\ni H\"afele, J.C. \& Keating, R.E. 1972 \emph{Science} \bf{177}, 166
 \newline\ni Hicken, M. \emph{et al}, 2009 \emph{Astro. J.} \bf{700}, 1097  table 1
 \newline\ni Kowalski, M. \emph{et al.} 2008 \emph{Astro. J.} \bf{686}, 749 
 \newline\ni Liddle, A.  2003  \emph{An Introduction to Modern Cosmology}, Wiley, Chichester, p.99
 \newline\ni Riess, A.G. \emph{et al.} 2009 \emph{Astro. J.} \bf{699}, 539 
 \newline\ni Sandage, A., Tammann, G.A., Saha, A., Reindl, B., Macchetto, F.D. \& Panagia, N. 2006 \emph{Astro. J.} \bf{653}, 843 

\sub{Legends}

Figure 1:  Supernova distance $r$ vs time with light path to earth AE.

\ni Figure 2:  Observed SN1A magnitudes M vs redshift z compared with equation (6) for $H_0=62.9\ km/s/Mpc$

\ni Figure 3: Residuals (observation - theory) vs magnitude M for $H_0=62.9\ km/s/Mpc$

\sub{Short title}
Does gravity operate between galaxies?

\begin{figure}[p]
\begin{center}
\resizebox{18cm}{!}{ \includegraphics{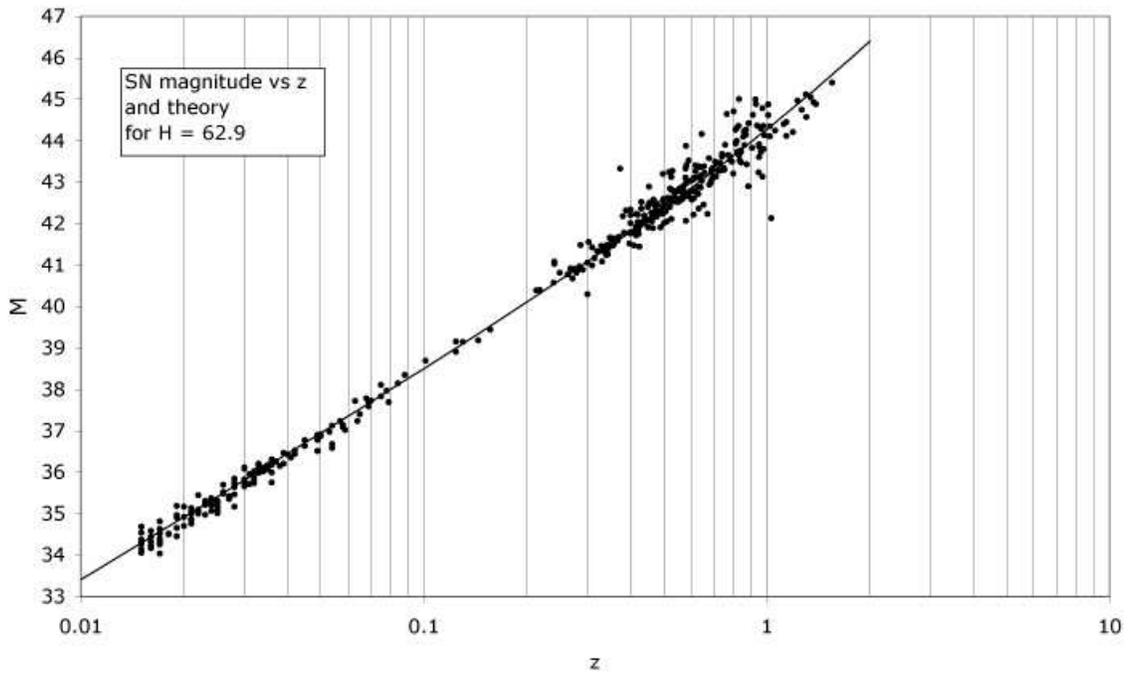}}
\caption{Observed SN1A magnitudes M vs redshift z compared with equation (6) for $H_0=62.9\ km/s/Mpc $}
\label{f2}
\end{center}
\end{figure}

\begin{figure}[p]
\begin{center}
\rotatebox{90}{\resizebox{11cm}{!}{ \includegraphics{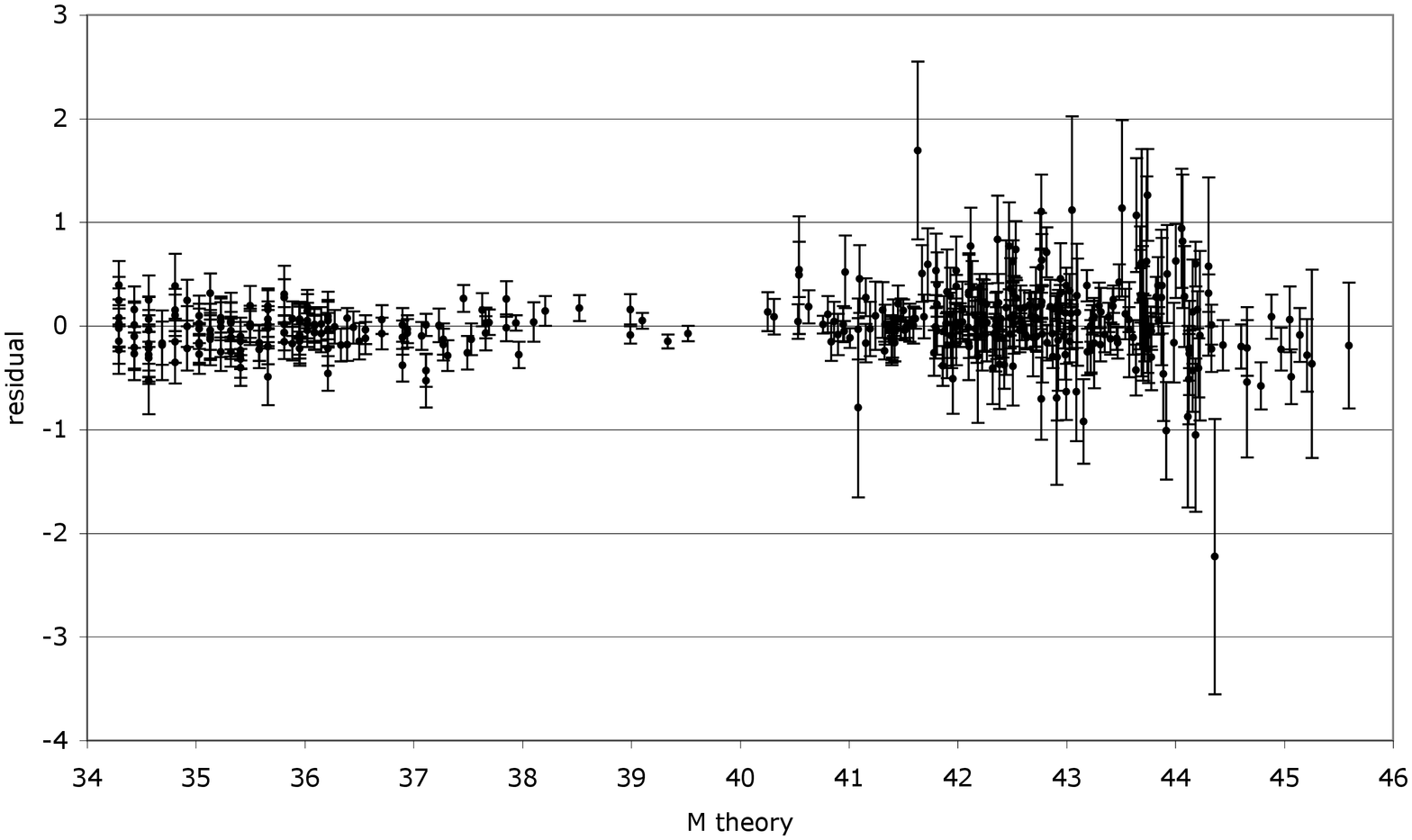}}}
\caption{Residuals (observation - theory) vs magnitude M for $H_0=62.9\ km/s/Mpc $}
\label{f3}
\end{center}
\end{figure}

\end{document}